**Title:** Uptake and outcome of manuscripts in Nature journals by review model and author characteristics

**Author list:** Barbara McGillivray* [1,2], Elisa De Ranieri [3]

[1] The Alan Turing Institute

[2] Theoretical and Applied Linguistics, Faculty of Modern and Medieval Languages, University of Cambridge

[3] Springer Nature

bmcgillivray@turing.ac.uk

e.deranieri@nature.com

**Abstract**

**Background**

Double-blind peer review has been proposed as a possible solution to avoid implicit referee bias in academic publishing. The aims of this study are to analyse the demographics of corresponding authors choosing double blind peer review, and to identify differences in the editorial outcome of manuscripts depending on their review model.

**Methods**

Data includes 128,454 manuscripts received between March 2015 and February 2017 by 25 Nature-branded journals. We investigated the uptake of double blind review in relation to journal tier and gender, country, and institutional prestige of the corresponding author. We then studied the manuscripts' editorial outcome in relation to review model and author's characteristics. The gender (male, female or NA) of the corresponding authors was determined from their first name using a third-party service (Gender API). The prestige of corresponding author's institutions was measured by using the Global Research Identifier Database (GRID) and dividing institutions in three prestige groups using the 2016 Times Higher Education (THE) ranking. We used descriptive statistics for data exploration and we tested our hypotheses using Pearson's chi-square and binomial tests.

**Results**

Author uptake for double-blind was 12%. We found a small but significant association between journal tier and review type. We had gender information for 50,533 corresponding authors, and found no statistically significant difference in the distribution of peer review model between males and females. We had 58,920 records with normalized institutions and a THE rank, and we found that corresponding authors from the less prestigious institutions are more likely to choose double-blind review. In the ten countries with the highest number of submissions, we found a small but significant association between country and review type. The outcome at both first decision and post review is significantly more negative (i.e. a higher likelihood for rejection) for double than single-blind papers.

**Conclusions**

Authors choose double-blind review more frequently when they submit to more prestigious journals, they are affiliated with less prestigious institutions or they are from specific countries; the double-blind option is also linked to less successful editorial outcomes.

# Background

Double-blind peer review (DBPR) has been proposed as a means to avoid (at least in principle) implicit bias from peer reviewers against characteristics of authors such as gender, country of origin or institution. Whereas in the more conventional single-blind peer review (SBPR) the reviewers have knowledge of the authors' identity and affiliations (Brown, 2006), in DBPR the identity and affiliations of the authors are hidden from the reviewers and vice versa. In spite of the presence of explicit instructions to authors, this type of review model has sometimes been shown to fail to hide authors' identity. Katz et al. report that 34% of 880 manuscripts submitted to two radiology journals contained information that would either potentially or definitely reveal the identities of the authors or their institution. Falagas et al. (2006) report that the authors of 47% of 614 submissions to the American Journal of Public Health were in fact recognizable.

Over the past years, several studies have analyzed the efficacy of DBPR in eradicating implicit bias in specific scientific disciplines. As detailed below, some of this work has shown evidence for gender and/or institution bias (cf. e.g. [4]), while other research has not found conclusive results (cf. e.g. [5]), demonstrating the need for further large-scale systematic analyses spanning over journals across the disciplinary spectrum ([6]; [7]; [8]). A recent study ([9]) with a controlled experiment found that in single-blind reviewers are more likely than double-blind reviewers to accept manuscripts from famous authors and high-ranked institutions.

[2] provide a systematic review and meta-analysis of biomedical journals and investigate the interventions aimed at improving the quality of peer review in these publications. The authors report that DBPR "did not affect the quality of the peer review report or rejection rate". [3] report similar results for the journal *Plastic and Reconstructive Surgery*.

Regarding gender bias, [1] showed that blinding interviewees in orchestra interviews led to more females being hired. In the context of scientific literature, [10] analyzed 2,680 manuscripts from 7 journals and found no overall difference in the acceptance rates of papers according to gender, while at the same time reporting a strong effect of number of authors and country of affiliation on manuscripts' acceptance rates. [11] analysed the distribution of gender among reviewers and editors of the Frontiers journals, showing an underrepresentation of women in the process, as well as a same-gender preference (homophily). [12] analyzed the journal *Behavioral Ecology*, which switched to DBPR in 2001; they found a significant interaction between gender and time, reflecting the higher number of female authors after 2001, but no significant interaction between gender and review type. [13] analyzed 940 papers submitted to an international conference on economics held in Sweden in 2008; they found no significant difference between the grades of female- and male-authored papers by review type. On the other hand, [14] analyzed EvoLang 11 papers and found that female authors received higher rankings under DBPR. Among the studies dealing with institution bias, [15] analyzed abstracts submitted to the American Heart Association's annual Scientific Sessions research meeting from 2000 to 2004, finding some evidence of bias favouring authors from English-speaking countries and prestigious institutions.

The study reported on here is the first one that focuses on Nature-branded journals, with the overall aim to investigate whether there any implicit bias in peer review in these journals, and ultimately understand whether DBPR is an effective measure in removing referee bias and improving the peer review of scientific literature. We focus on the Nature journals as the portfolio covers a wide range of disciplines in the natural sciences and biomedical research, and thus it gives us an opportunity to identify trends beyond discipline-specific patterns. In addition, the high prestige of these journals might accentuate an implicit referee bias and therefore makes such journals a good starting point for such analysis.

Nature-branded journals publishing primary research introduced DBPR as an optional service in March 2015 in response to authors' requests (Nature, no author listed). Authors have to indicate at first submission whether they wish to have their manuscript considered under SBPR or DBPR, and this choice is maintained if the manuscript is declined by one journal and transferred to another. If authors choose DBPR, their details (names and affiliations) are removed from the manuscript files, and it is the authors' responsibility to ensure their own anonymity throughout the text and beyond (e.g., authors opting for DBPR should not post on preprint archives). Editors are always aware of the identity of the authors.

In this study we sought to understand the demographics of authors choosing DBPR in Nature-branded journals and to identify any differences in success outcomes for manuscripts undergoing different review models and depending on the gender and the affiliation of the corresponding author. This study provides insight on author's behaviour when submitting to high-impact journals, and shows that manuscripts submitted under DBPR option are less likely to be sent to review and accepted than those submitted under SBPR.

## Methods

The study was designed to analyze the manuscripts submitted to Nature-branded journals publishing primary research between March 2015 (when the Nature-branded primary research journals introduced DBPR as an opt-in service) and February 2017.

For each manuscript we used Springer Nature's internal manuscript tracking system to extract name, institutional affiliation, and country of the corresponding author, journal title, the manuscript's review type (single-blind or double-blind), the editor's final decision on the manuscript (accept, reject, or revise), and the DOI. The corresponding author is defined here as the one author who is responsible for managing the submission process on the manuscript tracking system, but may not coincide with the corresponding author(s) as stated on the manuscript.

The dataset consisted of 133,465 unique records, with 63,552 different corresponding authors and 209,057 different institution names. In order to reduce the variability in the institutional affiliations, we normalized the institution names and countries via a Python script that queried the API of the Global Resource Identified Database (GRID [23]). We only retained a normalized institution name and country when the query to the GRID API returned a result with a high confidence, and the flag "manual review" was set to false, meaning that no manual review was needed. This process left 13,542 manuscripts without a normalized name; for the rest of the manuscripts, normalized institution names and countries were found, which resulted in 5,029 unique institution names.

In order to assign a measure of institutional prestige to each manuscript, we used the 2016/2017 Times Higher Education rankings (THE [20]), and normalized the institution names using the GRID API. We then mapped the normalized institution names from our dataset to the normalized institution names of the THE rankings via a Python script.

Finally, we associated each author with a gender label (male/female) by using the Gender API service [21].

The final dataset was further processed and then analysed statistically using the statistical programming language R, version 3.4.0. In the processing step we excluded 5,011 records which had an empty value in the column recording the review type due to technical issues in the submissions system for Nature Communications. These records are excluded from the analysis, resulting in a dataset of 128,454 records, of which 20,406 were submitted to Nature, 65,234 to the 23 sister journals and 42,814 to Nature Communications.

The dataset contains both direct submissions and transfers, i.e. manuscripts originally submitted to a journal and subsequently transferred to another journal which was deemed a better fit by the editor. In the case of transfers, the author cannot change the review type compared to the original submission, and therefore we excluded the 22,081 transferred manuscripts from the analysis of author uptake. We however included transfers in all other analyses because the analysed items are combinations of three attributes: paper, corresponding author and journal to which the paper was submitted.

We inspected the gender assigned via the Gender API, which assigns an accuracy score between 0 and 100 to each record. After manually checking a sample of gender assignments and their scores, we kept the gender returned by Gender API where the accuracy was at least 80, and assigned a value "NA" otherwise. This resulted in 17,379 instances of manuscripts whose corresponding author was female, 83,830 with male corresponding author, and 27,245 with gender NA. In the following analysis we will refer to the data where the gender field is not NA as the Gender Dataset.

Concerning the institutions, we defined four categories according to their THE ranks and used these as a proxy for prestige: category 1 included institutions with THE rank between 1 and 10, category 2 was for THE ranks between 11 and 100, category 3 for THE ranks above 100, and category 4 for non-ranked institutions. This choice of categories is arbitrary, e.g. we could have chosen a different distribution of institutions among the four categories, and will likely have an impact on the uptake of DBPR across the institutional prestige spectrum. However, we find that a logarithmic-based categorization of this sort would be more representative than a linear-based one. In the following analysis we will refer to the data for groups 1, 2 and 3 as the Institution Dataset.

We employed hypothesis testing techniques to test various hypotheses against the data. In order to test whether two variables were independent, we used Pearson's Chi-square test of independence and referred to the classification in [22] to define the strength of association. In order to test whether the proportions in different groups were the same we used the test of equal proportions in R (command "prop.test"). We used the significance threshold of 0.05.

# Results

We analysed the dataset of 128,454 records with a non-empty review type to answer the following questions:

1. What is the demographics of authors that choose double-blind peer review?

2. Which proportion of papers are sent for review under either SBPR or DBPR? Are there differences related to gender or institution within the same review model?

3. Which proportion of papers are accepted for publication under either SBPR or DBPR? Are there differences related to gender or institution within the same review model?

To place the results below within the right context, we point out that this study suffered from a key limitation, namely that we did not have an independent measure of quality for the manuscript, or a controlled experiment in which the same manuscript is reviewed under both peer review models. As a consequence, we are unable to distinguish bias towards author characteristics or the review model from any quality effect, and thus we can't conclude on the efficacy of DBPR to address bias. We discuss limitations of the study in more details in the Discussion section.

## 1. Analysis of peer review model uptake

We first analysed the demographics of corresponding authors that choose DBPR by journal group, gender, country and institution group. For this analysis, we only included direct submissions (106,373) and we excluded manuscripts that were rejected by one journal and then transferred to another. This is because authors cannot modify their choice of review model at the transfer stage, and thus transfers cannot contribute to the uptake analysis.

The overall uptake of DBPR is 12%, corresponding to 12,631 manuscripts, while for 93,742 manuscripts the authors chose the single-blind option.

Each of the 106,373 manuscripts was submitted to one of the 25 Nature-branded journals, and we investigated any potential differences in uptake depending on the journal tier. We divided the journals in three tiers: (i) the flagship interdisciplinary journal (Nature), (ii) the discipline-specific sister journals (Nature Astronomy, Nature Biomedical Engineering, Nature Biotechnology, Nature Cell Biology, Nature Chemical Biology, Nature Chemistry, Nature Climate Change, Nature Ecology & Evolution, Nature Energy, Nature Genetics, Nature Geoscience, Nature Human Behaviour, Nature Immunology, Nature Materials, Nature Medicine, Nature Methods, Nature Microbiology, Nature Nanotechnology, Nature Neuroscience, Nature Photonics, Nature Physics, Nature Plants, Nature Structural & Molecular Biology) and (iii) the open-access interdisciplinary title (Nature Communications).

Table 1 displays the uptake by journal group, and shows that the review model distribution changes as a function of the journal tier, with the proportion of double-blind papers decreasing for tiers with comparatively higher perceived prestige. We found a small but significant association between journal tier and review type. The results of a Pearson's chi-square test of independence are as follows: chi-squared = 378.17, degrees of freedom = 2, p-value<2.2e-16; Cramer's V = 0.054 and show that authors submitting to more prestigious journals tend to have a slight preference for DBPR compared to SBPR. This might indicate that authors are more likely to choose DB when the stakes are higher in an attempt to increase their success chances by removing any implicit bias from the referees.

**Table 1:** Uptake of peer review models by journal tier.

| Journal | DBPR | SBPR |
|---|---|---|
| Nature | 2,782 (14%) | 17,624 (86%) |
| Sister journals | 8,053 (12%) | 57,181 (88%) |
| Nature Communications | 3,900 (9%) | 38,914 (91%) |

We then analysed the uptake by gender as shown in Table 2, as we were interested in finding any gender-related patterns. For this analysis, we only considered 83,256 (out of the 106,373) manuscripts for which the gender assigned to the corresponding author's name by Gender API had a confidence score of at least 80 (the Gender Dataset, excluding transfers).

**Table 2:** Uptake of peer review models by gender of the corresponding author.

| Gender | DBPR | SBPR |
|---|---|---|
| Female | 1,506 (10%) | 12,943 (90%) |
| Male | 7,271 (11%) | 61,536 (89%) |

We did not find a significant association between gender and review type (Pearson's Chi-square test results: X-squared = 0.24883, df = 1, p-value = 0.6179), and thus we cannot reject the null hypothesis that gender and review type are independent.

In order to see if institutional prestige played a role in the choice of review type by authors, we analysed the uptake by institution group. For this analysis we used a subset of the 106,373 manuscripts consisting of 58,920 records with non-empty normalized institutions for which a THE rank was available (the Institution Dataset, excluding transfers). Table 3 shows that the proportion of authors choosing double-blind changes as a function of the institution group, with higher ranking groups having a higher proportion of single-blind manuscripts. This may be interpreted as indicating that authors from less prestigious institutions prefer the DB model as it prevents implicit referee bias against institutional affiliation.

**Table 3**. Uptake of peer review models by institution group.

| Institution group | DBPR | SBPR |
|---|---|---|
| | Actual | Actual |
| 1 | 240 (4%) | 5,818 (96%) |
| 2 | 1,663 (8%) | 19,295 (92%) |
| 3 | 4,174 (13%) | 27,730 (87%) |

We investigated the relationship between review type and institutional prestige (as measured by the institution groups) by testing the null hypothesis that the review type is independent from prestige. A Pearson's Chi-square test found a significant, but small association between institution group and review type (X-squared = 656.95, df = 2, p-value < 2.2e-16, Cramer's V = 0.106). We can conclude that authors from the least prestigious institutions are more likely to choose DBPR compared to authors from the most prestigious institutions and authors from the mid-range institutions.

Finally, we investigated the uptake of the peer review models by country of the corresponding author, using data on all of the 106,373 manuscripts. We found that 10 countries contributed to 80% of all submissions, and thus we grouped all other countries under the category 'Other'. Results on the uptake are shown in Table 4.

**Table 4**: Uptake of peer review model by country.

| Country | DBPR | SBPR |
|---------|------|------|
| Australia | 274 (10%) | 2366 (90%) |
| Canada | 259 (9%) | 2581 (91%) |
| China | 3,626 (22%) | 13148 (78%) |
| France | 278 (8%) | 3334 (92%) |
| Germany | 350 (5%) | 6079 (95%) |
| India | 711 (32%) | 1483 (68%) |
| Japan | 933 (15%) | 5248 (85%) |
| South Korea | 643 (12%) | 3,089 (88%) |
| United Kingdom | 509 (7%) | 6,656 (93%) |
| United States | 2,298 (7%) | 30,184 (93%) |
| Other | 2,750 (12%) | 19,574 (88%) |

Using Pearson's Chi-square test of independence, we found a small but significant association between country category and review type (X-squared = 3784.5, df = 10, p-value < 2.2e-16; Cramer's V = 0.189). Figure 1 shows a Cohen-Friendly association plot indicating deviations from independence of rows (countries) and columns (peer review model) in Table 4. The area of each rectangle is proportional to the difference between observed and expected frequencies, where the dotted lines refer to expected frequencies. The height of the rectangles is related to the significance and the width to the amount of data that support the result. China and US stand out for their strong preference for DBPR and SBPR, respectively. This might indicate that authors from countries with a more recent history of academic excellence are more concerned about implicit bias from referees against country.

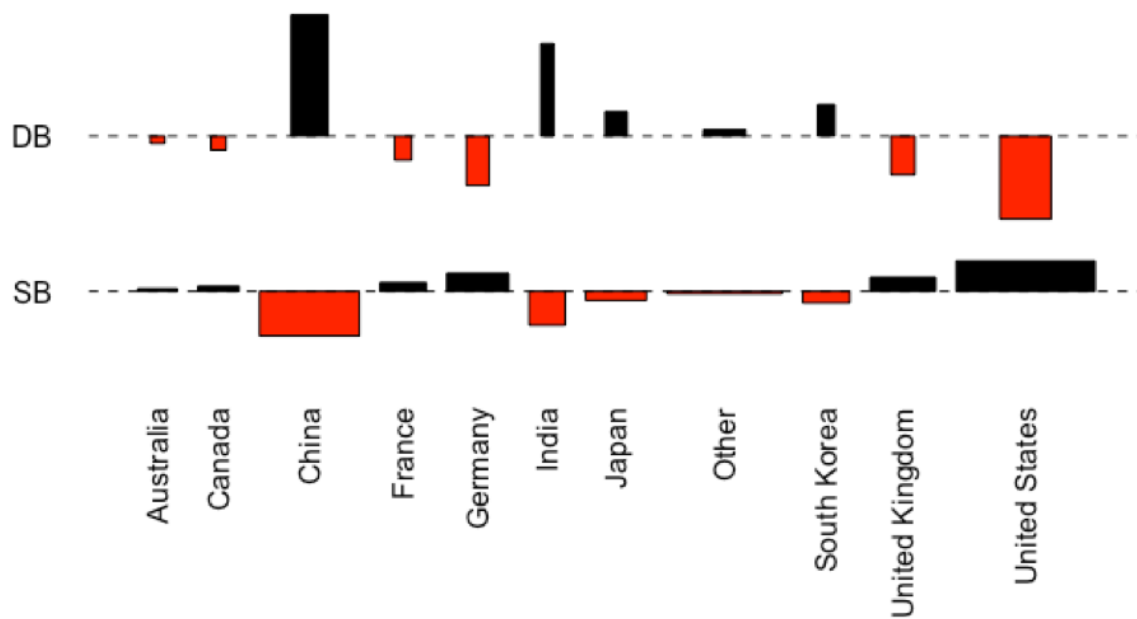

**Figure 1**: Cohen-Friendly association plot for Table 4.

## 2. Analysis at the out-to-review stage

Once a paper is submitted, the journal editors proceed with their assessment of the work and decide whether each manuscript is sent out for review (OTR) to external reviewers. This decision is taken solely by the editors, and they are aware of the chosen peer review model as well as all author information. We investigated the proportion of OTR papers (OTR rate) under both peer review models to see if there were any differences related to gender or institution. For this analysis we included direct submissions as well as transferred manuscripts, because the editorial criteria vary by journal and a manuscript rejected by one journal and transferred to another may then be sent out to review. Thus, our unit of analysis is identified by three elements: the manuscript, the corresponding author, and the journal.

Table 5 shows the counts and proportions of manuscripts that were sent out for review or rejected by the editors as a function of peer review model.

**Table 5:** Outcome of the first editorial decision (OTR rate) for papers submitted under the two peer review models.

| Outcome | DBPR | SBPR |
|---|---|---|
| Rejected outright | 13,493 (92%) | 87,734 (77%) |
| Out to review | 1,242 (8%) | 25,985 (23%) |

We found that a smaller proportion of DB papers are sent to review compared with SB papers, and that there is a very small but significant association between review type and outcome of the first editorial decision (results of a Chi-square test: X-squared = 1623.3, df = 1, p-value < 2.2e-16; Cramer's V = 0.112).  This can be explained by either an editor bias towards the review model, or by the fact that manuscripts submitted under DBPR are of a lower quality compared to SBPR manuscripts, or a combination of the two.

We also analysed the OTR rates by gender of the corresponding author, regardless of review type. Here, we included data on direct submissions and transfers (101,209 submissions). We excluded data where the gender was not assigned to either male or female. Results are in Table 6.

**Table 6:** Outcome of the first editorial decision (OTR rate) as a function of corresponding author's gender, and regardless of peer review model.

| Outcome | Female corresponding | Male corresponding |
|---|---|---|

| | authors | authors |
|---|---|---|
| Rejected outright | 13,493 (77.6%) | 65,046 (77.6%) |
| Out to review | 3,886 (22.4%) | 18,784 (22.4%) |

We did not find a significant association between OTR and gender (Pearson's Chi-square test results: X-squared = 0.015641, df = 1, p-value = 0.9005), so we cannot reject the null hypothesis that there is no association between OTR and gender. Our main question concerns a possible gender bias, therefore we investigated the relation between OTR rates, review model and gender, still including both direct submissions and transfers (Table 7).

**Table 7:** Outcome of the first editorial decision (OTR rate) as a function of corresponding author's gender and peer review model.

| Outcome | Female corresponding authors | | Male corresponding authors | |
|---|---|---|---|---|
| | DBPR | SBPR | DBPR | SBPR |
| Rejected outright | 1,549 (89.6%) | 11,944 (76.3%) | 7,835 (91.7%) | 57,211 (76.0%) |
| Out to review | 180 (10.4%) | 3,706 (23.7%) | 713 (8.3%) | 18,071 (24.0%) |

It seems that SB manuscripts by female corresponding authors are more likely to be rejected at the first editorial decision stage than those by male corresponding authors, and that DB manuscripts by male corresponding authors are less likely to be sent to review than those by female corresponding authors.

We tested the null hypothesis that the two populations (manuscripts by male corresponding authors and manuscripts by female corresponding authors) have the same OTR rate within each of the two review models. For this, we used a test for equality of proportions with continuity correction. For DB papers, we found a statistically significant difference in the OTR rate by gender (X-squared = 7.5042, df = 1, p-value = 0.006155); for SB papers we did not find a statistically significant difference in the OTR rate by gender (X-squared = 0.72863, df = 1, p-value = 0.3933). Therefore, in the DB case we can reject the null hypothesis and conclude that there is a significant difference between the OTR rate of papers by male corresponding authors and the OTR rate of papers by female corresponding authors. In the SB case, we cannot reject the null hypothesis. This can be explained by either an editor bias towards the review model, or the fact that female authors select their best papers to be DB to increase their chances of being accepted, or a combination of both.

Next, we focused on a potential institutional bias and looked at the relationship between OTR rate and institutional prestige as measured by the groups defined based on THE ranking explained above (excluding the fourth group, for which no THE ranking was available), regardless of review type (Table 8).

**Table 8:** Outcome of the first editorial decision (OTR rate) as a function of the group of the corresponding author's institution, regardless of peer review model.

| Outcome | Institution group 1 | Institution group 2 | Institution group 3 |
|---|---|---|---|
| Rejected outright | 4,541 (63%) | 18,949 (75%) | 32,046 (83%) |
| Out to review | 2,626 (37%) | 6,396 (25%) | 6,726 (17%) |

Papers from more prestigious institutions are more likely to be sent to review than papers from less prestigious institutions, regardless of review type. This is a statistically significant result, with a small effect size; the results of Pearson's chi-square test of independence are: X-squared = 1533.9, df = 2, p-value < 2.2e-16, Cramer's V = 0.147. This may be due to the higher quality of the papers from more prestigious institutions, or to an editor bias towards institutional prestige, or both.

Next, we investigated the relation between OTR rates, review model and institution group (Table 9), to detect any bias.

**Table 9:** Outcome of the first editorial decision (OTR rate) as a function of the group of the corresponding author's institution and the peer review model.

| Outcome | Institution group 1 | | Institution group 2 | | Institution group 3 | |
|---|---|---|---|---|---|---|
| | DBPR | SBPR | DBPR | SBPR | DBPR | SBPR |
| Rejected outright | 241 (85.8%) | 4,300 (62.4%) | 1,696 (86.8%) | 17,253 (73.8%) | 4,487 (91.8%) | 27,559 (81.3%) |
| Out to review | 40 (14.2%) | 2,586 (37.6%) | 259 (13.2%) | 6,137 (26.2%) | 399 (8.2%) | 6,327 (18.7%) |

We observe a trend in which the OTR rate for both DB and SB papers decreases as the prestige of the institution groups decreases, and we tested for the significance of this. A test for equality of proportions for group 1 and 2 for DB papers showed a non-significant result (X-squared = 0.13012, df = 1, p-value = 0.7183), and the same test on group 2 and group 3 for DB papers showed a significant result (X-squared = 40.898, df = 1, p-value = 1.604e-10). A test for equality of proportions for group 1 and 2 for SB papers returned a significant difference (X-squared = 331.62, df = 1, p-value < 2.2e-16); the same test for group 2 and group 3 for SB papers also returned a significant difference (X-squared = 464.86, df = 1, p-value < 2.2e-16).

Across the three prestige group, SBPR papers are more likely to be sent to review. This can be explained by either an editor bias towards the review model, or a self-selection effect in which authors within each institution group choose to submit their best studies under SBPR, or a combination of both.

### 3. Analysis of outcome post-review

Finally, we investigated the outcome of post-review decisions as a function of peer review model and characteristics of the corresponding author, to try and identify patterns of potential referee bias. We studied whether papers were accepted or rejected following peer review, and we included transfers because the editorial decisions at different journals follow different criteria. We excluded papers for which the post-review outcome was a revision and papers which were still under review, thus the dataset for this analysis comprises 20,706 records of which 8,934 were accepted and 11,772 were rejected. The decision post-review of whether to accept a paper or not is taken by the editor but is based on the feedback received from the referees, so we assume that the decision at this stage would reflect a potential referee bias.

Table 10 displays the accept rate by review type defined as the number of accepted papers over the total number of accepted or rejected papers.

**Table 10:** Acceptance rate by review type.

| Outcome | DBPR | SBPR |
|---|---|---|
|  |  |  |
| Accepted | 242 (25%) | 8,692 (44%) |
| Rejected | 732 (75%) | 11,040 (56%) |

We found that DB papers that are sent to review have an acceptance rate that is significantly lower than that of SB papers. The results of a Pearson's Chi-square test of independence show a small effect size (X-squared = 138.77, df = 1, p-value < 2.2e-16; Cramer's V = 0.082). Two possible explanations for the observed differences are a referee bias against DB, and the lower quality of the studies submitted under double-blind peer review. These results, which are in line with the analysis at the OTR stage described in section 2, can be explained by either a lower quality of manuscripts submitted under DBPR, or by an implicit bias of referees towards the review model, or both.

We investigated the question of whether, out of the papers that go to review, manuscripts by female corresponding authors are more likely to be accepted than those with male corresponding authors under DBPR and SBPR. We excluded the records for which the assigned gender was NA, and focused on a dataset of 17,167 records of which 2,849 (17%) had a female corresponding author and 14,318 (83%) had a male corresponding author.

First, we calculated the acceptance rate by gender, regardless of review type (Table 11).

**Table 11**: Outcome for papers sent to review as a function of the gender of the corresponding author, regardless of review model.

| Outcome | Female corresponding authors | Male corresponding authors |
|---|---|---|
| | | |
| Accepted | 1,222 (43%) | 6,434 (45%) |
| Rejected | 1,627 (57%) | 7,884 (55%) |

We observe a significant but very small difference in the acceptance rate by gender (Pearson's Chi-square test of independence: X-squared = 3.9364, df = 1, p-value = 0.047; Cramer's V = 0.015), leading us to conclude that manuscripts by female corresponding authors are slightly less likely to be accepted. We identify two potential causes for this, one being a difference in quality and the other being a gender bias.

To ascertain whether indeed any referee bias is present, we studied the acceptance rate by gender and review type. Table 12 shows the proportion of manuscripts that are sent for review and accepted or rejected with different peer review model and by gender of the corresponding author.

**Table 12.** Outcome of papers sent to review by gender of the corresponding author and by review model.

| Outcome | Female corresponding authors | | Male corresponding authors | |
|---|---|---|---|---|
| | DBPR | SBPR | DBPR | SBPR |
| Accepted | 35 (26%) | 1,187 (44%) | 137 (25%) | 6,297 (46%) |
| Rejected | 99 (74%) | 1,528 (56%) | 413 (75%) | 7,471 (54%) |

If we compare the proportion of accepted manuscripts under DB and authored by female vs. male corresponding authors (26% vs. 25%) with a test for equality of proportions with continuity correction, we find that there is a not significant difference female authors and male authors for DB accepted papers (results of 2-sample test for equality of proportions with continuity correction: X-squared = 0.03188, df = 1, p-value = 0.8583).

If we compare male authors' and female authors' acceptance rates for SB papers (44% vs 46%), we find that there is not a significant difference female authors and male authors for SB accepted manuscripts (results of 2-sample test for equality of proportions with continuity correction test: X-squared = 3.6388, df = 1, p-value = 0.05645).

Based on these results, we cannot conclude whether the referees are biased towards gender.

In order to detect any bias towards institutional prestige, we focused on the three institution groups we defined (high, medium, and low-prestige), thus excluding the fourth group for which no THE rank was found; the dataset contained 12,054 records, which includes OTR papers that were either rejected or accepted, as well as transfers. Table 13 shows acceptance rate by institution group, regardless of review type.

**Table 13.** Outcome of manuscripts sent to review as a function of the institution group of the corresponding author, regardless of review model.

| Outcome | Institution group 1 | Institution group 2 | Institution group 3 |
|---------|---------------------|---------------------|---------------------|
| Accepted | 996 (49%) | 2,108 (43%) | 2,078 (40%) |
| Rejected | 1,029 (51%) | 2,743 (57%) | 3,100 (60%) |

There is a tiny but significant association between institution group and acceptance (Pearson's Chi-square test results: X-squared = 49.405, df = 2, p-value = 1.87e-11, Cramer's V = 0.064), which means that authors from less prestigious institutions tend to be rejected more than authors from more prestigious institutions, regardless of review type. The difference, however, is very small. This can be due to quality or referee bias, so we tested this by looking at differences in dependence of the review model.

**Table 14.** Outcome of manuscripts sent to review as a function of the institution group of the corresponding author and review model.

| Outcome | Institution group 1 | | Institution group 2 | | Institution group 3 | |
|---|---|---|---|---|---|---|
| | DBPR | SBPR | DBPR | SBPR | DBPR | SBPR |
| Accepted | 11 (37%) | 985 (49%) | 61 (30%) | 2,047 (44%) | 75 (23%) | 2,003 (41%) |
| Rejected | 19 (63%) | 1,010 (51%) | 139 (70%) | 2,604 (56%) | 251 (77%) | 2,849 (59%) |

In order to measure any quality effect, we tested the null hypothesis that the populations (institution group 1, 2, and 3) have the same proportion of accepted manuscripts for DB manuscripts with a test for equality of proportions (proportion of accepted manuscripts: 0.37 for group 1, 0.31 for group 2, and 0.23 for group 3). The test yielded a non-significant p-value (X-squared = 5.2848, df = 2, p-value = 0.07119), meaning that we cannot say that there is a significant difference between authors from prestigious institutions and authors from less prestigious institutions for DB accepted manuscripts.

We tested the null hypothesis that the populations (institution group 1, 2, and 3) have the same proportion of accepted manuscripts for SB manuscripts with a test for equality of proportions (proportion of accepted manuscripts: 0.49 for group 1, 0.44 for group 2, and 0.41 for group 3).

We found a significant result (X-squared = 37.76, df = 2, p-value = 6.318e-09). This means that there is a statistically significant difference between the three groups. In order to identify the pair(s) giving rise to this difference, we performed a test of equal proportion for each pair, and accounted for multiple testing with Bonferroni correction. The results were significant for all pairs: group 1 vs. group 2 (X-squared = 15.961, df = 1, p-value = 6.465e-05*3=0.0001939603); group 2 vs. group 3 (X-squared = 7.1264, df = 1, p-value = 0.007596*3 = 0.02278758); group 1 vs. group 3 (X-squared = 37.304, df = 1, p-value = 1.011e-09*3=3.031627e-09). However, we were still unable to find the reason for this effect conclusively. This result may occur as a consequence of various scenarios. On the one hand, the could be a referee bias towards institution groups, which leads to manuscripts from more prestigious institutions to be treated more favourably. On the other hand, the manuscripts from more prestigious institutions may be of higher quality. Finally, the authors may select their best papers to be reviewed via SBPR.

# Discussion

DBPR was introduced in the Nature journal in response to the author community's wish for a bias-free peer review process. The underlying research question that drove this study is to assess whether DBPR is effective in removing or reducing implicit bias in peer review. However (as mentioned above and discussed below in more detail) the fact that we did not control for the quality of the manuscripts means that the conclusions on the efficacy of DBPR that can be drawn from this data is limited. Any conclusive statement about the efficacy of DBPR would have to wait until such control can be implemented, or more data collected. Nevertheless, the available data allowed us to draw conclusions on the uptake of the review models, as we detail below.

The results on author uptake show that DBPR is chosen more frequently when authors submit to higher impact journals within the portfolio, when they are from specific countries (India and China in particular, among countries with the highest submission rates) and when they are from less prestigious institutions. However, no difference by corresponding author's gender was found. One interpretation of these results is that authors chose DBPR when there is increased perceived risk of discrimination, with the exception of gender discrimination. Thus, authors that feel more vulnerable to implicit bias against the prestige of their institution or their country tend to choose DBPR to prevent such bias playing a role.

In addition, authors might choose one review model over another depending on the quality of the work. We could hypothesize that authors choose to submit their best works as SBPR as they are proud of it, and lesser quality work as DBPR; however, the opposite hypothesis could be true, i.e. that authors prefer to submit their best work as DBPR to give it a fairer chance against implicit bias. Either hypothesis may apply to a different demographics of authors, and moreover we were unable to independently measure the quality of the manuscripts so this "self-selection" effect on the author's part remains undetermined in our study.

The analysis of success outcome at both the out-to-review and accept stages shows that DBPR manuscripts are less likely to be successful than SBPR manuscripts (Tables 5 and 10). This can be due to either bias of editors and reviewers towards review model, or to differences in the quality of the manuscripts. As editors know the author's information, any editor's bias towards gender, country or institutional prestige would also play a role in the out-to-review stage in particular. For example, there are more DBPR papers that are from China, and if editors were biased against that country this would reflect in the OTR rate for DBPR papers.

In order to distinguish bias from any quality effect we would need an independent measure of quality (e.g. citations in the case of published articles), or a controlled experiment in which the same paper is reviewed under both models. Alternatively, a dataset where DBPR is compulsory would eliminate the effect of bias towards the review model. Considering citations as a proxy for quality, besides providing an imperfect picture, also limits the size of the dataset to published articles that have had time to accrue citations. In our case, such dataset would have been very small, given the low acceptance rate of the journals considered, and the fact that the dataset is very recent (due to the date in which DBPR was introduced in Nature journals). Moreover, a controlled experiment, or a mandate of DBPR for all submissions, was not possible due to peer review policies at the Nature journals and the fact that we analysed historic data.

Thus our discussion is limited to what observed in this dataset, and our analysis could not disentangle the effects of bias towards author characteristics, of bias towards the review model, and of quality of the manuscripts. From our experience, we are inclined to think that the Nature journals editors are not biased, which seems to be corroborated by the fact (Table 6) that there is no significant association between gender and OTR rate, regardless of peer review model, for example. This suggests that there is no editor bias towards gender. If editors were truly unbiased, we could conclude that the quality of DBPR papers is inferior to that of SBPR ones. This is an unproven assumption, however this interpretation may be supported also by the results (Tables 8 and 9) on the OTR rate by institution group, as it is not unthinkable to assume that (on average) manuscripts from more prestigious institutions, which tend to have better funding more resources, are of a higher quality than those from institutions with lesser prestige and means.

When comparing acceptance rates by gender and regardless of review model, we observed that female authors are less likely to be accepted than their male counterparts, and this is a significant difference. However, we were unable to distinguish the effects of gender bias and manuscript quality in this observation because an analysis of accept rate by gender and review type did not yield statistically significant results.

Finally, the post-review outcome of papers as a function of the institution group and review model (Table 14) showed that manuscripts from less prestigious institutions are accepted less than those from more prestigious ones even under DBPR, however due to the small numbers of papers at this stage the results are not statistically significant.

We should note that we did not perform any multivariate analysis, and the significance of the main results on outcome is limited by the size of the dataset for accepted papers, due to the high selectivity of these journals and to the low uptake of DBPR. We calculated that, at this rate, it would take us several decades to collect sufficient data that would result in statistically significant results, so another strategy is required, e.g. making DBPR compulsory to accelerate data collection and remove potential bias against review model. Also, because of the retrospective nature of this study, we could not conduct controlled experiments. In future works, we will consider studying the post-decision outcome also in relation to the gender of reviewers, and defining a quality metrics for manuscripts in order to isolate the effect of bias.

## Conclusions

This study is the first one that analyzes and compares the uptake and outcome of manuscripts submitted to scientific journals covering a wide range of disciplines depending on the review model chosen by the author (double-blind vs. single-blind peer review). We have analysed a large dataset of submissions to 25 Nature journals over a period of 2 years by review model and in dependence of characteristics of the corresponding author. Our aim was to understand the demographics of author uptake, and infer the presence of any potential implicit bias towards gender, country, or institutional prestige in relation to the corresponding author.

We observed that authors tend to choose DBPR more often when submitting to higher impact journals within the Nature portfolio, when they are from specific countries (India and China in particular, among countries with the highest submission rates) and when they are from less prestigious institutions. We did not observe any difference by gender.

Because of the small size of the data set for accepted papers we did not find a significant conclusion on whether any implicit bias towards gender or institutional prestige exists, however we did see that manuscripts from female authors or authors from less prestigious institutions are accepted with a lower rate than those from male authors or more prestigious institutions, respectively.

## List of abbreviations

DBPR: double-blind peer review

GRID: Global Research Identifier Database

OTR: out to review

SBPR: single-blind peer review

THE: Times Higher Education

## Ethics approval and consent to participate

Not applicable.

## Consent for publication

Not applicable

# Availability of data and materials

The data that support the findings of this study are available from Springer Nature, but restrictions apply to the availability of these data, which were used under license for the current study, and so are not publicly available. Data are however available from the authors upon reasonable request and with permission of Springer Nature.

# Competing interests

EDR is employed by Macmillan Publishers Ldt., which publishes the Nature-branded journals.

# Funding

This work was supported by The Alan Turing Institute under the EPSRC grant EP/N510129/1.

# Authors' contributions

BMcG collected the data from GRID and THE, processed the data, and conducted the statistical analysis. BMcG was the major contributor in writing the Background and Methods sections. EDR proposed the study and provided the data on manuscript submissions and the gender data from Gender API. EDR was the major contributor in writing the Discussion and Conclusions sections. Both authors designed the study and contributed equally to the Results section. Both authors read and approved the final manuscript.

# Acknowledgements

We would like to thank Michelle Samarasinghe for help in collecting the data from the manuscript tracking system and Sowmya Swaminathan for comments on the study design and feedback on the manuscript draft.